**Penalized Poisson model for network meta-analysis of individual patient time-to-event data**

OLLIER Edouard[1,2,3], Blanchard Pierre[2,4], LE TEUFF Gwénaël[1,2*], MICHIELS Stefan[1,2*]

[1]Service de Biostatistique et d'Épidémiologie, Gustave Roussy, Université Paris-Saclay, Villejuif, France

[2]Oncostat U1018, Inserm, Université Paris-Saclay, labeled Ligue Contre le Cancer, Villejuif, France

[3]SAINBIOSE U1059, Equipe DVH, Université Jean Monnet, Saint-Etienne, France

[4] Département de radiothérapie, Gustave Roussy, Université Paris-Saclay, Villejuif, France

*Co-last authors

Corresponding author:
Edouard Ollier
Unité de Recherche Clinique, Innovation, Pharmacologie
Hôpital Nord, CHU de Saint-Etienne, 42055 Saint- Etienne Cedex 2
E-Mail: edouard.ollier@univ-st-etienne.fr

**Abstract**

Network meta-analysis (NMA) allows the combination of direct and indirect evidence from a set of randomized clinical trials. Performing NMA using individual patient data (IPD) is considered as a "gold standard" approach as it provides several advantages over NMA based on aggregate data. For example, it allows to perform advanced modelling of covariates or covariate-treatment interactions. An important issue in IPD NMA is the selection of influential parameters among terms that account for inconsistency, covariates, covariate-by-treatment interactions or non-proportionality of treatments effect for time to event data. This issue has not been deeply studied in the literature yet and in particular not for time-to-event data. A major difficulty is to jointly account for between-trial heterogeneity which could have a major influence on the selection process. The use of penalized generalized mixed effect model is a solution, but existing implementations have several shortcomings and an important computational cost that precludes their use for complex IPD NMA. In this article, we propose a penalized Poisson regression model to perform IPD NMA of time-to-event data. It is based only on fixed effect parameters which improve its computational cost over the use of random effects. It could be easily implemented using existing penalized regression package. Computer code is shared for implementation. The methods were applied on simulated data to illustrate the importance to take into account between trial heterogeneity during the selection procedure. Finally, it was applied to an IPD NMA of overall survival of chemotherapy and radiotherapy in nasopharyngeal carcinoma.

Key words: network meta-analysis, individual patient data, time-to-event data, penalized likelihood, lasso, frequentist, model selection.

**Introduction**

Network meta-analyses (NMA) allow the combination of direct and indirect evidence from a set of randomized clinical trials to estimate all pairwise treatment comparisons. Performing NMA using individual patient data (IPD) is considered as a "gold standard" approach as it provides the most accurate estimates of treatments effect and the greatest power to identify relevant covariates or covariate-treatment interactions[1-2]. NMA can be performed directly on the whole individual patient dataset (one stage approach) or using summarized estimates of treatments effect obtained separately by applying a standardized statistical procedure to IPD of each study (two stages approach).

IPD NMA has been well studied for continuous or binary outcomes[3-4] but fewer works have been conducted for time-to-event data. Recently, a one-stage Bayesian approach based on the Royston-Parmar model have been proposed[5], resulting in a very flexible model. From the frequentist point of view, any methods estimating treatment effect such as the Cox proportional hazard model or the Peto estimator could be used to perform a two-stage IPD NMA[6]. However, in the presence of between-trial heterogeneity and/or non-proportionality of treatment effect, the use of this approach may be suboptimal and one-stage approach are preferable[3-4]. For standard IPD meta-analysis, the use of a Poisson generalized mixed linear model have been proposed[7] and allows to easily perform one stage meta-analysis while accounting for heterogeneity and non-proportionality. Furthermore, this approach relies on the framework of generalized linear models and then benefit from a low computational burden. To our knowledge, this approach has not been yet applied to non-trivial IPD NMA for time-to-event data with more than 3 treatments.

Generally, realization of an IPD NMA faces two key issues. The first one is the quantification of between-trial heterogeneity in treatment effects as for standard MA. Accounting for between-trial heterogeneity becomes mandatory when heterogeneity is large and remain unexplained in order to avoid bias in treatment effect estimations. This can be done by using mixed effects models consisting on inclusion of random effects. Compared to fixed effect models, estimation of mixed effects models requires to solve a complex optimization problem as the likelihood corresponds to a multidimensional integral that is not known explicitly. This is particularly true for IPD NMA, where multiple random effects may be included in the statistical model, making inference particularly complex especially with time-to-event data. The use of Bayesian inference techniques may be a solution as it does not require the use of complex optimization procedure and may explain why NMA commonly use Bayesian modelling.

The second issue is the selection of the network meta-analysis models. Indeed, in NMA several additional parameters could be included in the model. Inconsistency hypothesis violation could be taken into account by including parameters for treatment loops in which inconsistency is present[8]. Covariates and covariate-by-treatment interactions could be included in the model to account for the association between baseline hazard or treatment effect with some patients' characteristics[9]. With time-to-event data, non-proportionality of treatment effects could also be modelled by including interaction between treatment and follow-up time[5]. All these parameters could be included as fixed regression coefficients, but a selection step has to be performed to define which parameters has to be included. Even if the issue of variables selection is not a high dimensional one, the size of the set of all possible models grows rapidly with the number of parameters which depends on the geometry of the network defined by the number of treatments (nodes) and connection between treatments (edges). The model building process may then benefit from an automatic selection routine. Several methods such as stepwise procedures and penalized likelihood approaches are classically used in this context of variables selection and may be directly applied to IPD NMA. Recently, a simulation study[10] suggested that penalized likelihood methods may be more suitable for parameter selection in IPD meta-analysis.

In IPD NMA, the method used to analyze the data need then ideally to perform model selection and jointly take into account between-trials heterogeneity. This may be done by using penalized Generalized Linear Mixed Model (GLMM)[11-12]. However, such models are particularly difficult to estimate as their likelihood is both not known explicitly and non-differentiable. Several algorithms have been proposed[11-13], but existing implementations have multiple shortcomings. The main one is the impossibility to jointly estimate penalized and unpenalized parameters which is necessary in order not to bias treatments effect estimations. An alternative solution would be to develop a fixed effect model that accounts for between-trial heterogeneity. This could be done by implementing a penalized model in which study specific random effects are treated as fixed effects but penalized in order to ensure identifiability. This would thus increase the number of parameters in the model. This approach has already been used to analyze stratified data with lasso penalty where each parameter is decomposed into a sum of a common effect plus a stratum-specific parameter[14].

In this work, we propose a Poisson regression model to perform IPD network meta-analysis of time-to-event data using a frequentist approach. Heterogeneity quantification and model selection effect are performed simultaneously by using a penalized fixed effect model allowing to overcome the optimization problem met when applying random effect models. The proposed method can be easily implemented using existing penalized regression package and R code is shared on https://github.com/EdOllier/PenalizedPoissonNMA. Section 2 introduces a NMA in head and neck; section 3 describes our statistical penalized method; section 4 presents the simulation framework; section 5 presents the results of the simulations and the application of our approach to a real example.

**NMA of chemotherapy and radiotherapy in nasopharyngeal carcinoma**

The IPD NMA of chemotherapy and radiotherapy in nasopharyngeal carcinoma aimed to investigate the benefit of different timing of chemotherapy in addition to a radiotherapy (RT) on overall survival[6]. The network includes 7 treatments (nodes): RT alone (reference treatment), induction chemotherapy (IC) followed by RT (IC-RT); RT followed by adjuvant chemotherapy (RT-AC); IC followed by RT followed by AC (IC-RT-AC); chemo-radiotherapy (CRT); IC followed by CRT (IC-CRT); and CRT followed by AC (CRT-AC). The network is composed of 4806 individual patient data from 19 trials split into 24 comparisons (edges). The number of trials per comparison varied from 1 to 7. The structure of the network is represented in Figure 1. The original analysis of overall survival data was based on a two-stage fixed effect-model. Treatment effects (hazard ratios) were estimated separately for each trial based on IPD in a first step using the Peto estimator and then combined using a NMA in a second step. Heterogeneity, inconsistency, and non-proportionality of treatments effect were investigated. No covariates effects or covariate-by-treatment interactions were investigated at the time.

**Poisson regression model for network meta-analysis**

The Poisson regression model has been previously proposed for standard meta-analysis of time-to-event IPD[7] and may be directly extended to NMA of IPD. In this section, we present its fixed treatment effects extension in the context of NMA of individual patient time-to-event data studying Q different treatments (the first treatment (q=1) is set as reference):

$$d_{ijk} \sim \mathcal{P}(\mu_{ijk})$$

$$log(\mu_{ijk}) = log(\xi_{ijk}) + \pi_k \times 1_{k \neq 1} + \gamma_i + \sum_{q=2}^{Q} \beta_q \times trt_{ij}^q$$

where $d_{ijk}$ corresponds to the event indicator of patient j (j=1,…,$n_i$) in trial i (i=1,…,N) during period k (k=1,…,K). It takes the value of 0 or 1. It is modeled by a Poisson process for each patient of each trial during each time period. The constant $log(\xi_{ijk})$ is the time at risk of patient j in trial i during period k and is included as an offset in the model. The baseline hazard function is treated as a piecewise constant function over K time periods. Parameters $\pi_k$ ($\pi \in R^{K-1}$) correspond to the adjustment for the k-th time period (k=2,…,K). Parameters $\gamma_i$ ($\gamma \in R^N$) correspond to the study specific baseline hazard rate of study i during the first period (k=1).

Variables $trt_{ij}^q$ are the treatment contrast comparing treatment q to a fixed reference treatment and can take the values 0, 1 or -1:

$$trt_{ij}^q = \begin{cases} 1 & \text{if the patient is randomized to treatment q as experimental treatment} \\ -1 & \text{if patient is randomized to the experimental arm with treatment q as reference} \\ 0 & \text{otherwise} \end{cases}$$

For example, in a trial i comparing the q-th treatment versus the p-th treatment, treatment contrast variables take the following values for a patient j in the experimental arm:

$$trt_{ij}^q = 1$$
$$trt_{ij}^p = -1$$
$$trt_{ij}^l = 0 \ (l \neq p \ and \ l \neq q).$$

For patients in the reference arm, all treatment contrasts are set to 0.

Finally, parameter $\beta_q$ corresponds to the log hazard ratio for the q-th treatment versus the reference treatment. Under the consistency hypothesis, the log hazard ratio can be reconstructed for every possible pairwise treatment comparison:

$$\log HR_{q \ vs \ p} = \beta_q - \beta_p$$

To decrease computation time of Poisson modelling for IPD NMA, it is possible to collapse individual data across treatments, periods and covariates patterns, with $d_{ijk}$ corresponding then to the number of events in the i-th study during the k-th period in subjects belonging the j-th pattern of covariates (for example : male of the experimental arm). The same Poisson regression model can be fitted to this collapsed dataset giving exactly the same parameter estimates as the Poisson regression model fitted to the non-collapsed dataset. This process is however limited to the case where all covariates are categorical.

**Penalized Poisson model and between-trial heterogeneity**

Between-trial heterogeneity of treatment effect for a particular comparison is usually taken into account by including a random effect into the statistical model for each treatment comparison. When multiple treatment comparisons (edges) are included in the network, the inference will quickly become very complex with multiple integrations. We propose a penalized fixed effects model as an alternative to this inference issue.

For a NMA, random effects can be included in the Poisson regression previously described in the following manner:

$$log(\mu_{ijk}) = log(\xi_{ijk}) + \pi_k \times 1_{k \neq 1} + \gamma_i + \sum_{q=2}^{Q} \beta_{qi} \times trt_{ij}^q \quad (1)$$

$$\gamma_i = \bar{\gamma} + u_i \ with \ u_i \sim \mathcal{N}(0, \sigma^2)$$
$$\beta_{qi} = \bar{\beta}_q + v_{qi} \ with \ v_{qi} \sim \mathcal{N}(0, \tau_q^2)$$

Where $\bar{\gamma} \epsilon R$ and $\bar{\beta} \epsilon R^{Q-1}$ correspond to the average baseline hazard and relative treatment effects. Hidden variables $(u_i)_{i=1,\ldots,N}$ and $(v_{qi})_{i=1,\ldots,N}$ correspond to study specific random effects. Parameters $\sigma^2$ and $(\tau_q^2)_{q=2,\ldots,Q}$ correspond to the between-trial variance for baseline hazard and relative treatment effects, respectively.

The vector of fixed effect parameter $\theta = (\pi, \bar{\gamma}, \bar{\beta})$ and $\sigma^2$ and $(\tau_q^2)_{q=2,...,Q}$ could be estimated by solving the following maximum likelihood problem:

$$\underset{\theta, \sigma^2, \tau_q^2}{\operatorname{argmin}} \left\{ -\log \iint P(d, u, v) \, du \, dv \right\}$$

where $P(d, u, v)$ represents the complete data likelihood written as follows:

$$-\log P(d, u, v) \propto -\log P(d|\theta, u, v) - \log P(u, v|\theta)$$

$$\propto -\log P(d|\theta, u, v) + \frac{N}{2} \log(\sigma^2) + \frac{1}{2\sigma^2} \| u \|_2^2 + \sum_q \left( \frac{N}{2} \log(\tau_q^2) + \frac{1}{2\tau_q^2} \| v_q \|_2^2 \right)$$

with $P(d|\theta, u, v)$ the Poisson likelihood of the observation given $\theta, u$ and $v$

$$\log P(d|\theta, u, v) = \sum_{i,j,k} \left( d_{ijk} \times \log(\mu_{ijk}) + \mu_{ijk} - \log(d_{ijk}!) \right)$$

and $\|\cdot\|_2^2$ is the Euclidian norm.

This mixed Poisson model (equation 1) can be used to derive a penalized Poisson regression problem in which hidden variables $u$ and $v$ are treated as fixed effects.

$$\log(\mu_{ijk}) = \log(\xi_{ijk}) + \pi_k \times 1_{k \neq 1} + (\bar{\gamma} + u_i) + \sum_{q=2}^{Q} (\bar{\beta}_q + v_{qi}) \times trt_{ij}^q \quad (2)$$

$$\underset{\theta, u, v_q}{\operatorname{argmin}} \left\{ -\log P(d|\theta, u, v) + \frac{\lambda_0}{2} \| u \|_2^2 + \sum_q \frac{\lambda_q}{2} \| v_q \|_2^2 \right\} \quad (3)$$

where $\lambda_0$ and $(\lambda_q)_{q=2,...,Q}$ are the penalization parameters. These penalization parameters are linked to between-trial variances of random effects:

$$\lambda_0 = \frac{1}{\sigma^2} \text{ and } \lambda_q = \frac{1}{\tau_q^2} \quad (q = 2, ..., Q) \quad (4)$$

Within this framework, estimation of $\theta, u$ and $v$ can be obtained easily by estimating a fixed penalized Poisson regression problem which is computationally less intensive than solving equation 1. Additionally, it is possible to fit a model with a common between-trial variance for all the relative treatment effects by setting the corresponding penalization parameters to a common value ($\lambda_q = \lambda \in R$ for all q).

Penalization parameters can be calibrated[15] by solving iteratively equation (3) and updating values $\lambda_0$ and $(\lambda_q)_{q=2,...,Q}$:

$$\begin{pmatrix} \theta_{n+1} \\ u_{n+1} \\ (v_{q,n+1})_{q=2,\ldots,Q} \end{pmatrix} = \underset{\theta,u,v_q}{\operatorname{argmin}} \left\{ -\log P(d|\theta,u,v) + \frac{\lambda_0^n}{2} \| u \|_2^2 + \sum_q \frac{\lambda_q^n}{2} \| v_q \|_2^2 \right\}$$

$$\lambda_0^n = \frac{df_{u_n}}{u_n^t u_n}$$

$$\lambda_q^n = \frac{df_{v_{q,n}}}{(v_{q,n})^t v_{q,n}}$$

with $df_{u_n}$ and $df_{v_{q,n}}$ the respective degrees of freedom of $u_n$ and $v_{q,n}$

$$df_{u_n} = Trace\ (H_{u_n,u_n})$$
$$df_{v_{q,n}} = Trace\ (H_{v_{q,n},v_{q,n}})$$

The matrix H corresponds to the hat matrix and is defined as:

$$H = \left[ \frac{\partial^2 \log P(d|\Theta)}{\partial \Theta^t \partial \Theta} - diag(\lambda) \right]^{-1} \left[ \frac{\partial^2 \log P(d|\Theta)}{\partial \Theta^t \partial \Theta} \right]$$

with $\Theta = (\theta, u, v)$ and $\lambda = (0, \ldots, 0, \lambda_0, \lambda_{q=2,\ldots,Q})$ the vector of regularization parameter for each component of $\Theta$. The Q+K first component of vector $\lambda$ are 0 as parameters in $\theta$ ($\theta \in R^{Q+K}$) are not penalized. In the case of Poisson regression, the hessian matrix is:

$$\frac{\partial^2 \log P(d|\Theta)}{\partial \Theta^t \partial \Theta} = X^t diag(\mu) X$$

With $d$ and $\mu$ the vectors of events indicators and Poisson rates, respectively. The matrix $X$ corresponds to the design matrix of the linear predictor:

$$d \sim \mathcal{P}(\mu)$$
$$log(\mu) = log(\xi) + X\Theta$$

**Parameters selection with adaptive lasso penalty**

In IPD NMA, treatment effects do not have to undergo the selection process and are then always included in the model. But some parameters accounting for inconsistency, covariates or covariate-treatment interactions effects and non-proportionality of treatment effects do not have to be systematically included in the model. The goal of the modelling process is then to perform variable selection on these effects.

The use of the penalized Poisson regression framework presented before allows to include L[1] norm penalty to perform variable selection[16]. Therefore, several terms can be added in the Poisson regression model respectively to take into account:

i) Inconsistency

Inconsistency corresponds to a discrepancy between the treatment effect estimates based on direct and indirect comparisons. In a treatment loop between three treatments A, B, and C, an inconsistency is observed if

$$log(\widehat{HR}_{B\ vs\ C}^{Direct}) \neq log(\widehat{HR}_{B\ vs\ C}^{Indirect}) = \beta_C - \beta_B$$

Where $\widehat{HR}_{B\ vs\ C}^{Direct}$ is the estimate of the effect of treatment B compared to treatment C considering only trials that directly compared B and C. To take into account inconsistency[8], we may introduce an inconsistency parameter $\omega_{BC}$ such as

$$log(\widehat{HR}_{B\ vs\ C}^{Direct}) = log(\widehat{HR}_{B\ vs\ C}^{Indirect}) + \omega_{BC} = \beta_C - \beta_B + \omega_{BC}$$

To quantify this inconsistency, the following term is introduced in the Poisson regression model:

$$\sum_{1<q<p} \omega_{qp} \times |trt_{ij}^q \times trt_{ij}^p|$$

The term $|trt_{ij}^q \times trt_{ij}^p|$ being equal to 1 in patients randomized to the experimental arm of trials that directly compared the q-th and the p-th treatments and equal to 0 otherwise. Parameter $\omega_{qp}$ quantifies then the discrepancy between direct and indirect comparisons in the loop defined by the q-th, the p-th and the reference treatment. It is included in the model, if and only if a loop between the q-th, the p-th and the reference treatment is present within the treatments network.

ii) Covariate effects

$$\sum_{c=1}^{C} \delta_c \times z_{ijc}$$

With $z_{ijc}$ the values of the c-th covariates for the j-th patient of the i-th trial and $\delta_c$ the parameter that quantify the effect of the c-th covariates on baseline hazard.

iii) Covariate-by-treatment interactions

$$\sum_{q=2}^{Q} \sum_{c=1}^{C} \alpha_{cq} \times z_{ijc} \times trt_{ij}^q$$

With $\alpha_{cq}$ the parameter that quantify the effect of the interaction between the c-th covariates and the q-th treatment.

iv) Non-proportionality of treatments effects

The non-proportionality of treatment effects is modelled as a time interval by treatment interaction with time divided into K intervals considering the first time interval (k=1) as the reference:

$$\sum_{q=2}^{Q}\sum_{k=2}^{K} \zeta_{kq} \times trt_{ij}^q \times 1_{t \in \mathcal{P}_k}$$

With $\zeta_{kq}$ the parameter that quantify the effect of the interaction between the k-th follow-up period and the q-th treatment and $1_{t \in \mathcal{P}_k}$ the indicator variable of the k-th time period.

When all these terms are added to the Poisson regression model (2), we obtain the following model:

$$log(\mu_{ijk}) = log(\xi_{ijk}) + \pi_k \times 1_{k \neq 1} + (\bar{\gamma} + u_i) + \sum_{q=2}^{Q}(\bar{\beta}_q + v_{qi}) \times trt_{ij}^q$$

$$+ \sum_{1<q<p} \omega_{qp} \times \left|trt_{ij}^q \times trt_{ij}^p\right| + \sum_{c=1}^{C} \delta_c \times z_{ijc} + \sum_{q=2}^{Q}\sum_{c=1}^{C} \alpha_{cq} \times z_{ijc} \times trt_{ij}^q$$

$$+ \sum_{q=2}^{Q}\sum_{k=2}^{K} \zeta_{kq} \times trt_{ij}^q \times 1_{t \in \mathcal{P}_k}$$

As parameters $\omega$, $\delta$, $\alpha$ and $\zeta$ have to be selected, we penalize the likelihood (3) using an adaptive Lasso penalty[17] (weighted $L_1$ norm) on these parameters:

$$\underset{\theta,u,v_q}{\mathrm{argmin}} \left\{ \begin{array}{l} -\log P(d|\theta,u,v) + \frac{\lambda_0}{2} \| u \|_2^2 + \sum_q \frac{\lambda_q}{2} \| v_q \|_2^2 \\ + \lambda_{L_1}\left(\| \rho_{\hat{\omega}} \circ \omega \|_1 + \| \rho_{\hat{\delta}} \circ \delta \|_1 + \| \rho_{\hat{\alpha}} \circ \alpha \|_1 + \| \rho_{\hat{\zeta}} \circ \zeta \|_1\right) \end{array} \right\} \quad (5),$$

with ∘ the Hadamard product and $\lambda_{L_1}$ the coefficient that tune the degree of sparsity in the final estimates of $\omega$, $\delta$, $\alpha$ and $\zeta$. The vector of fixed effects is now $\theta = (\pi, \bar{\gamma}, \bar{\beta}, \omega, \delta, \alpha, \zeta)$. Penalty weights $\rho_\omega$, $\rho_\delta$, $\rho_\alpha$ and $\rho_\zeta$ of the adaptive lasso penalty can be computed using unpenalized estimates of $\omega$, $\delta$, $\alpha$ and $\zeta$ obtain with $\lambda_{L_1}$ equal or close to 0:

$$\rho_\omega = \frac{1}{\hat{\omega}_{UnPen}}, \rho_\delta = \frac{1}{\hat{\delta}_{UnPen}}, \rho_\alpha = \frac{1}{\hat{\alpha}_{UnPen}} \text{ and } \rho_\zeta = \frac{1}{\hat{\zeta}_{UnPen}}.$$

For a fixed value of $\lambda_{L_1}$, penalization parameters $\lambda_0$ and $(\lambda_q)_{q=2,\ldots,Q}$ are calibrated by solving iteratively equation 5 with updated values $\lambda_0$ and $(\lambda_q)_{q=2,\ldots,Q}$:

$$\begin{pmatrix} \theta_{n+1} \\ u_{n+1} \\ (v_{q,n+1})_{q=2,\dots,Q} \end{pmatrix} = \underset{\theta,u,v_q}{\mathrm{argmin}} \left\{ \begin{array}{l} -\log P(d|\theta,u,v) + \frac{\lambda_0^n}{2} \parallel u \parallel_2^2 + \sum_q \frac{\lambda_q^n}{2} \parallel v_q \parallel_2^2 \\ + \lambda_{L_1}\left( \parallel \rho_{\hat{\omega}} \circ \omega \parallel_1 + \parallel \rho_{\hat{\delta}} \circ \delta \parallel_1 + \parallel \rho_{\hat{\alpha}} \circ \alpha \parallel_1 + \parallel \rho_{\hat{\zeta}} \circ \zeta \parallel_1 \right) \end{array} \right\},$$

$$\lambda_0^n = \frac{df_{u_n}}{u_n{}^t u_n},$$

$$\lambda_q^n = \frac{df_{v_{q,n}}}{(v_{q,n})^t v_{q,n}}.$$

At each iteration, the solution of this penalized optimization problem could be solved using the R package penalized[20]. This package solves L1 and/or L2 penalized generalized regression models and allows to penalize each parameter with different types of penalties (L1, L2 or both).

Shrinkage coefficient $\lambda_{L_1}$ is selected using a two-step BIC approach[18]. The first step corresponds to the evaluation of the solution of equation (5) for a grid of user-defined $\lambda_{L_1}$ values. In a second step, the models selected with each $\lambda_{L_1}$ value of the grid are re-estimated without L1 penalty. The optimal $\lambda_{L_1}$ value is the one that minimizes the BIC calculated with a degree of freedom set to:

$$df = Trace\left(H_{S_{\hat{\Theta}}, S_{\hat{\Theta}}}\right),$$

with $S_{\hat{\Theta}}$ the support of the estimated vector of parameters $\hat{\Theta}=(\hat{\theta}, \hat{u}, \hat{v})$. The whole procedure was called HetAdLASSO in the rest of this work.

**Simulation framework**

We performed a simulation study to evaluate estimation accuracy and selection model performance of the penalized Poisson approach in the framework of IPD NMA for time-to-event data. Simulations were based on a non-trivial network of 5 treatments (A, B, C, D and E) with several loops. The geometry of the network is represented in figure 2 (5 nodes and 9 edges). This structure was chosen to avoid identifiability problem due to structure of the network and then avoid confusion between the performance of the proposed method and structural identifiability problems. Treatment A was considered as the reference treatment with direct comparison with treatments B, C, D, and E. IPD time-to-event data were simulated using the R package survsim[19]. Event and censoring times were both simulated using an exponential distribution with scale parameter set to -5.5 and -7 respectively (logarithmic scale). From these parameters, the average censoring rates was around 21%. For each pairwise comparison, different number of trials were considered (3, 5 and 10 trials). The sample size of each trial was drawn from a uniform distribution (U(50,500)). The coefficients associated to each treatment contrast variable were set to: $\bar{\beta}_{B\ vs\ A} = \log(0.77)$, $\bar{\beta}_{C\ vs\ A} = \log(0.65)$, $\bar{\beta}_{D\ vs\ A} = \log(0.96)$, $\bar{\beta}_{E\ vs\ A} = \log(0.87)$. These treatment effects correspond to moderate effect size. Treatment effects for other comparisons were derived from basic parameters and consistency hypothesis. Standard deviation of between-trial heterogeneity for baseline hazard was set to $\sigma = 0.2$. Standard deviation of between-trial heterogeneity $(\tau_q)_{q=B,\ldots,E}$ were considered equal for each comparison with the reference treatment and different values were considered (0.1, 0.2, 0.3, 0.4 and 0.5). We considered two binary covariates (C=2) simulated from a Bernoulli distribution with a probability of 0.5. The follow-up time was divided in K=6 periods. We simulated 5 scenarios with increasing complexity according to different model specifications (true model). These specifications represent different situations in terms of inconsistency, covariates, covariate-treatment interaction and non-proportionality parameters (Table 1). Scenario 1 represent a simple NMA with no inconsistency, covariate, covariate-treatment interaction and non-proportionality of treatment effects ($\omega = 0$, $\delta = 0$, $\alpha = 0$ and $\zeta = 0$). Scenario 2 represents a NMA with inconsistency in two treatments loops but without covariate, covariate-treatment interaction and non-proportionality of treatment effects. Inconsistency is simulated in treatments loops ABC and ADE through a hazard ratio (log scale) for BC and DE comparison different to the difference between BA and CA for loop ABC and to the difference between DA and EA for loop ADE, respectively. Scenario 3 represents a NMA with a covariate and two covariate-treatment interactions effects without inconsistency and non-proportionality of treatment effects. Scenario 4 represents a NMA with inconsistency, covariate and covariate-treatment interaction effects without non-proportionality. Scenario 5 represent a NMA with both inconsistency, covariate, covariate-by-treatment interactions and non-proportionality of treatment effects. Inconsistency, covariates and covariate-treatment interactions were parametrized in the same way as in scenario 4. Treatment E was considered to have a non-proportional treatment effect compared to reference treatment A. As the survsim package does not allow to implement directly non proportional treatment effects, we used a different baseline hazard for subject receiving treatment E compared to the other subjects. This trick allows to simulate the time-dependent of treatment effect for treatment E. In arms involving treatment E, a Weibull hazard with shape parameter set to 0.75 was used to simulate event times. The other arms were supposed to follow an exponential hazard as previously described. Hazard shapes used are shown in supplementary figure 1.

For each scenario and each between-trial heterogeneity levels, we simulated 100 data sets. We applied the proposed penalized adaptive lasso procedure (equation 5, HetAdLASSO) on each simulated data set. Between-trial heterogeneity was evaluated on baseline hazard and each treatments contrast. To assess the impact of between-trial heterogeneity on model selection performances, we also applied an adaptive lasso procedure that does not account for between-trial variability (FxAdLASSO for Fixed Adaptive Lasso). It is equivalent to solve HetAdLASSO described by equation (5) but without random effects, which consists in excluding from HetAdLASSO the terms associated to u and v.

Several criteria were used to evaluate the selection performances of HetAdLASSO and FxAdLASSO. First, to evaluate selection performance at the parameter level, false positive rate (FPR, proportion of non-influent parameters that are selected), false negative rate (FNR, proportion of influent parameters that are not selected) and accuracy (ACC, proportion of the model's parameters that were correctly identified within each data set) were calculated for each simulated data set:

$$FPR = \frac{FP}{TN + FP}, \quad FNR = \frac{FN}{FN + TP}, \quad ACC = \frac{TP + TN}{TP + TN + FN + FP}.$$

With i) TP the number of influent parameters that were selected, ii) TN the number of non-influent parameters that were not selected, iii) FP the number of non-influent parameters that were selected and iv) FN the number of influent parameters that were not selected.

We also evaluated the proportion of correctly selected model according to each specification of the true model (inconsistency, covariates, covariate-treatment interactions and non-proportionality). Lastly, we estimated the absolute bias of (i) each treatment contrast for the 2 approaches and (ii) the variance of the between-trial heterogeneity of treatment effects returned by HetAdLASSO. Both procedures (FxAdLASSO and HetAdLASSO) were implemented using the R package penalized[20] and our R code is available on the repository https://github.com/EdOllier/PenalizedPoissonNMA.

**RESULTS**

**Simulation study**

Figure 3 represents selection accuracy, FNR and FPR of both adaptive LASSO methods for the different simulated scenarios with increasing between-trial heterogeneity of treatment effects. Overall, selection accuracy of HetAdLASSO is close to 100% and does not seem to be influenced by the level of between-trial variance. When between-trial heterogeneity is not taken into account (FxAdLASSO), selection accuracy decreased for increasing between-trial variance. The decreasing selection accuracy of FxAdLASSO was mainly driven by an increased false positive rate. On the contrary, an increasing between-trial variance tends to increase the FNR of HetAdLASSO mainly driven by the non-detection of inconsistency terms.

Figure 4 represents the proportion of times both methods recovered the true model according to the parameter types (covariates, covariate-by-treatments interactions, inconsistency and non-proportionality) in the different simulated scenarios. Covariate parameters were correctly selected for both methods and do not seem to be affected by the level of between-trial variance. Concerning covariate-treatment interactions parameters, FxAdLASSO performances decreased slightly with between-trial heterogeneity of treatment effects. Major differences between FxAdLASSO and HetAdLASSO methods were observed for the selection of inconsistency and non-proportionality parameters. For scenario with moderate and high between-trial heterogeneity values, selection performance decreases mainly driven by the issue of inconsistency detection for both methods. FxAdLASSO selected falsely positive inconsistency terms whereas HetAdLASSO was more conservative and fails to detect significant inconsistency term (false negative). FxAdLASSO approach fails also to select the correct non-proportionality model with decreasing performance for increasing between-trial heterogeneity. This is not the case of HetAdLASSO approach. Overall, FxAdLASSO performed also well with low between-trial heterogeneity except for inconsistency characteristic, but selection accuracy decreases with increasing between-trial heterogeneity mainly driven by an increased false positive rate. Taking the between-trial heterogeneity into account with the HetAdLASSO significantly increased selection performance.

Figure 5 represents the box plot of absolute bias for log-hazard ratio $(\bar{\beta}_q)_{q=2,\dots,Q}$ parameters obtained with both methods. FxAdLASSO tends to return biased estimates when between-trial heterogeneity increased. Using HetAdLASSO allows to reduce this bias problem and its variability.

Finally, it is also possible to compare the true values of between-trial standard deviation ($(\tau_q)_{q=2,\dots,Q}$) to the estimated values from the penalization parameters ($(\lambda_q)_{q=2,\dots,Q}$) obtained using the iterative procedure (equation 4). These results are presented in figure 6. In these simulations, although it was not directly design for, the HetAdLASSO method allows a good estimation of between-trial standard deviations with increasing precision for an increasing number of trials.

**Application to IPD network meta-analysis of chemotherapy and radiotherapy in nasopharyngeal carcinoma**

We applied HetAdLASSO and FxAdLASSO methods to overall survival data of the IPD NMA in nasopharyngeal carcinoma. Due to the network's structure and the limited number of trials in some comparisons then: i) no between-trials heterogeneity was estimated for IC_RT_AC treatment contrast (only one trial evaluated IC_RT_AC) and ii) only one inconsistency parameter was estimated, in the loop defined by treatments CRT and CRT_AC. Adding inconsistency parameter in the other loops would lead to identifiability issues especially for HetAdLASSO.  Five covariates were included in the analysis: age (treated as a continuous covariate), sex (male (reference), female), Tumor size stage (T1 (reference), T2, T3 and T4), node invasion stage (N0 (reference), N1, N2 and N3) and radiotherapy technique (2D (reference), 2.5D, 3D and IMRT). Categorical covariates were converted to indicator variables. Follow-up time was divided in K=4 periods to guarantee a sufficient number of events in each time interval. Covariate-by-treatment interactions were investigated for each treatment contrast and covariates. This makes a total number of 84 parameters subject to the selection process and penalized by the lasso penalty.

Models respectively selected with HetAdLASSO and FxAdLASSO methods are summarized in Table 2. None of the methods selected covariate-by-treatment interactions and non-proportionality parameters. Only FxAdLASSO selected the parameter of inconsistency for the loop defined by treatments RT, CRT and CRT_AC. Concerning covariates, HetAdLASSO selected age, sex, tumor size stage T4 and node invasion stage N3 as influent covariates. FxAdLASSO selected 3 additional covariates: IMRT radiotherapy technique, tumor size stage T3 and node invasion stage N2. These results are coherent with the previous simulation study, in which FxAdLASSO tended to select more complex model. Table 3 summarized parameters estimates obtained with both methods. Treatments effect estimates of HetAdLASSO are coherent with the ones obtained with the two-stage analysis previously published[6] and the re-analyze of the available data (data from one of the two trials evaluating IC_RT_AC was not available because investigators did not allow the use of their data for this methodological research). Estimates of between-trials heterogeneity of treatment effect, obtained using equation 4, seems negligeable in all treatment (tau≤0.05) contrast except for CRT with tau=0.34. Based on the results of this network meta-analysis, we can conclude that chemoradiotherapy followed by adjuvant chemotherapy (CRT_AC) is the treatment that achieved the highest overall survival benefit in nasopharyngeal carcinoma.

**Discussion**

Network of meta-analysis of individual patient time-to-event data represents the state-of-the-art approach for evidence synthesis of time-to-event outcomes. However, conducting such an analysis necessitates to perform a model selection procedure to identify influent covariates, covariates-treatment interaction, inconsistency and non-proportionality parameters. The literature on the subject is relatively sparse. One of the major difficulties being taking into account the between-trials heterogeneity which could influence the selection process. This work presented a penalized Poisson model in which between trials differences are treated as fixed effects which allows to take into account between trials heterogeneity. The simulation study shows the advantage to take into account the between-trial heterogeneity compared to a fixed approach.

The proposed method faces some limitations. First, our method corresponds to a penalized regression approach and then does not return directly standard errors of parameters estimation and confidence intervals. In the application to nasopharyngeal carcinoma data, standard errors were computed for the selected model by performing a non-penalized re-estimation of the selected parameters on bootstrapped data sets. However, even if it is a direct and easy way to obtain an estimation of uncertainty in parameters estimate, it does not consider the uncertainty due to the selection procedure. It can lead to an underestimation of parameters estimation uncertainty. Several authors studied this post-selection inference problem and proposed different solutions[21-22], but these solutions are not trivial and are beyond the scope of this work.

Secondly, our method does not provide a direct estimation of between trial variances. Our approach corresponds to an empirical Bayesian method that rely on the empirical variance of the posterior mode of between trials differences (u and v). Based on estimated values of penalization parameters $\lambda_0$ and $(\lambda_q)_{q=2,...,Q}$, equation 4 could be used to return an approximation of between trial variances. For data sets with a large number of trials with large sample sizes, these values may be a good approximation as posterior distribution of between trials differences u and v are asymptotically normal[23]. This was the case in our simulation study. However, with fewer trials or more limited sample sizes posterior distributions of between trials differences may be skewed; the posterior mode and posterior mean of between trials differences may then differ significantly. In such a case, the empirical variance of posterior mode of u and v may provide a biased estimation of between trial variances. This phenomenon does not appear in the EM-algorithm as the calculation of between trial variances are based on conditional mean of u and v.

Last, our simulation study did not explore the impact of network structure on parameters estimation. We purposely used a densely connected network to avoid the identifiability problem due to the network structure. The latter could have a non-negligible impact on the results and although the effect of network topology on NMA results has been very little studied, some pioneering work has highlighted the impact of network structure irregularity on the bias and standard deviation of treatment effect estimates[24].

The present work may be extended in several ways. First, in this work no random effects were introduced on parameters $\pi_k$, however it may be of interest to quantify between trial variability for the time evolution of baseline hazard rate. These random effects could directly be implemented and estimated using the same procedure as the one used for parameter $\gamma$. Moreover, the evolution over time of the baseline hazard rate is defined as a piecewise constant function, with the number of time periods (K) being specified by the user. The selection of the appropriate value for K may be an issue. In the nasopharyngeal carcinoma example, K=4 time periods were chosen to guarantee a sufficient number of events in each time interval. In practice, 1-year interval are often used, as in the nasopharyngeal carcinoma meta-analysis[26]. A pragmatic alternative commonly used is to use the percentiles of the distribution of the events time. Murray proposed a formal calculation of the number of time interval K[27]. This calculation is based on the observed number of events and the percentiles of the events time. A possible extension of our work would be to consider a large number of time periods and penalized differences between successive parameters $\pi_k$ to regularize their estimation. This can be done by using ridge fusion penalty for example. This type of penalty may also be applied to treatment effects. Indeed, treatments effects were not penalized as it would shrink coefficient toward zero and then bias treatments effect estimates. However, in a selection perspective, it may be interesting to detect treatments that have the same effect. This would be possible by penalizing treatments effects using a fusion penalty that penalize difference between treatments effects[28,29] and then promote sparsity in the difference between parameters. The last extension concerning the baseline hazard rate would be to use continuous follow-up time; this can be achieved by replacing the proposed piecewise constant model by a spline basis[5] to take into account the possible nonlinear relationship between follow-up time and baseline hazard rate.

Finally, to explore the presence of ecological bias it has been proposed to decompose covariate-by-treatment interactions in a within and a between trial interactions[9,25]. We do not consider such a decomposition in our work, but this could be directly implemented in the proposed method. Moreover, specific parameters for between trial interactions interaction terms could also be penalized using a lasso penalty and then be included in the selection procedure.

**Conclusion**

To conclude, in this work we developed a penalized Poisson method to select network meta-analysis model based on individual patient time-to-event data. Our approach takes into account between trials heterogeneity without using the framework of mixed effect models that are computationally more intensive especially for large datasets encountered in meta-analysis of individual patient time-to-event data.

**Acknowledgement**
EO thanks the Nuovo Soldati Fellowship Fund for its support.

The authors would like to thank the MAC-NPC Collaborative Group (cf. Ribassin-Majed et al.[6] for the list of its members) who allowed to use the data for methodological research.

**Data Availability Statement**
The nasopharyngeal carcinoma data that support the findings of this study are not publicly available. Restrictions apply to the availability of these data. Data requests should be made to Gustave Roussy.

# References

1. Berlin JA, Santanna J, Schmid CH, Szczech LA, Feldman HI. Individual patient- versus group-level data meta-regressions for the investigation of treatment effect modifiers: ecological bias rears its ugly head. Stat Med. 2002;21:371-387.

2. Lambert PC, Sutton AJ, Abrams KR, Jones DR. A comparison of summary patient-level covariates in meta-regression with individual patient data meta-analysis. J Clin Epidemiol. 2002;55:86-94.

3. Debray T P, Schuit E, Efthimiou O et al. An overview of methods for network meta-analysis using individual participant data: when do benefits arise? Stat Methods Med Res. 2018;27(5):1351-64.

4. Debray TP, Moons KG, Abo-Zaid GM, Koffijberg H, Riley RD. Individual participant data meta-analysis for a binary outcome: one-stage or two-stage? PLoS One. 2013;8(4):e60650.

5. Freeman SC, Carpenter JR. Bayesian one-step IPD network meta-analysis of time-to-event data using Royston-Parmar models. Res Synth Methods. 2017;8(4):451-464.

6. Ribassin-Majed L, Marguet S, Lee AWM, Tong Ng W, Ma J, Chan ATC, Yu Huang P, Zhu G, Chua TT, Chen Y, Qiang Mai H, Kwong DLW, Lee Cheah S, Moon J, Tung Y, Kwan-Hwa C, Fountzilas G, Bourhis J, Pignon JP, Blanchard P. What Is the Best Treatment of Locally Advanced Nasopharyngeal Carcinoma? An Individual Patient Data Network Meta-Analysis. J Clin Oncol. 2017;35(5):498-505.

7. Crowther MJ, Riley RD, Staessen JA, Wang J, Gueyffier F, Lambert PC. Individual patient data meta-analysis of survival data using Poisson regression models. *BMC Med Res Methodol*. 2012;*12*(1):34.

8. Lu G, Ades AE. Assessing evidence inconsistency in mixed treatment comparisons. J Am Stat Assoc. 2006;101(474):447-459.

9. Freeman SC, Fisher D, Tierney JF, Carpenter JR. A framework for identifying treatment-covariate interactions in individual participant data network meta-analysis. Res Synth Methods. 2018;9(3):393-407.

10. Seo M, White IR, Furukawa TA, Imai H, Valgimigli M, Egger M, Zwahlen M, Efthimiou O. Comparing methods for estimating patient-specific treatment effects in individual patient data meta-analysis. Stat Med. 2021 Mar 15;40(6):1553-1573.

11. Groll A, Tutz G. Variable selection for generalized linear mixed models by L 1-penalized estimation. Stat Comput. 2014;24(2):137-154.

12. Schelldorfer J, Meier L, Bühlmann P. Glmmlasso: an algorithm for high-dimensional generalized linear mixed models using ℓ1-penalization. J Comput Graph Stat. 2014;23(2):460-477.

13. Fort G, Ollier E, Samson A. Stochastic proximal-gradient algorithms for penalized mixed models. Stat Comput. 2019;29(2):231-253.

14. Ollier E, Viallon V. Regression modelling on stratified data with the lasso. Biometrika, 2017;104(1):83-96.

15. Perperoglou A. Cox models with dynamic ridge penalties on time-varying effects of the covariates. Stat Med. 2014;33(1):170-180.

16. Tibshirani R. Regression shrinkage and selection via the lasso. J R Stat Soc Series B Stat Methodol. 1996;58(1):267-288.

17. Zou, H. The adaptive lasso and its oracle properties. J Am Stat Assoc. 2006;101(476):1418-1429.

18. Efron B, Hastie T, Johnstone I, Tibshirani R. Least angle regression. Ann Stat. 2004;32(2):407-499.

19. Moriña D, Navarro A. The R package survsim for the simulation of simple and complex survival data. J Stat Softw, 2014;59(2):1-20.

20. Goeman JJ. L1 penalized estimation in the Cox proportional hazards model. Biom J. 2010;52(1):70-84.

21. Lee JD, Sun DL, Sun Y, Taylor JE. Exact post-selection inference, with application to the lasso. Ann Stat. 2016;*44*(3), 907-927.

22. Taylor J, Tibshirani R J. Statistical learning and selective inference. *Proc Natl Acad Sci U S A*. 2015;*112*(25), 7629-7634.

23. Baghishani H, Mohammadzadeh M. Asymptotic normality of posterior distributions for generalized linear mixed models. J Multivar Anal. 2012;111:66-77.

24. Davies AL, Galla T (2021). Degree irregularity and rank probability bias in network meta-analysis. Res Synth Methods. 12(3), 316-332.

25. Hua H, Burke DL, Crowther MJ, Ensor J, Tudur Smith C, Riley RD. One-stage individual participant data meta-analysis models: estimation of treatment-covariate interactions must avoid ecological bias by separating out within-trial and across-trial information. Stat Med. 2017;*36*(5):772-789.

26. Blanchard P, Lee A, Marguet S, Leclercq J, Ng WT, Ma J, Chan AT, Huang PY, Benhamou E, Zhu G, Chua DT, Chen Y, Mai HQ, Kwong DL, Cheah SL, Moon J, Tung Y, Chi KH, Fountzilas G, Zhang L, Hui EP, Lu TX, Bourhis J, Pignon JP; MAC-NPC Collaborative Group.Lancet Oncol. 2015;16(6):645-55.

27. Murray TA, Hobbs BP, Lystig TC, Carlin BP. Semiparametric Bayesian commensurate survival model for post-market medical device surveillance with non-exchangeable historical data. Biometrics. 2014;70(1):185–191.

28. Ollier E, Samson A, Delavenne X, Viallon V. A SAEM algorithm for fused lasso penalized non-linear mixed effect models: application to group comparison in pharmacokinetics. Comput Stat Data Anal. 2016;95:207-221.

29. Viallon V, Lambert-Lacroix S, Höfling H, Picard F. On the robustness of the Generalized Fused Lasso to prior specifications. Stat Comput. 2016;26(1):285:301.

**Figure 1** Treatments network of the data set for chemotherapy timing in nasopharyngeal carcinoma. The number of trials (t) and number patients (pts) is shown for each comparison.

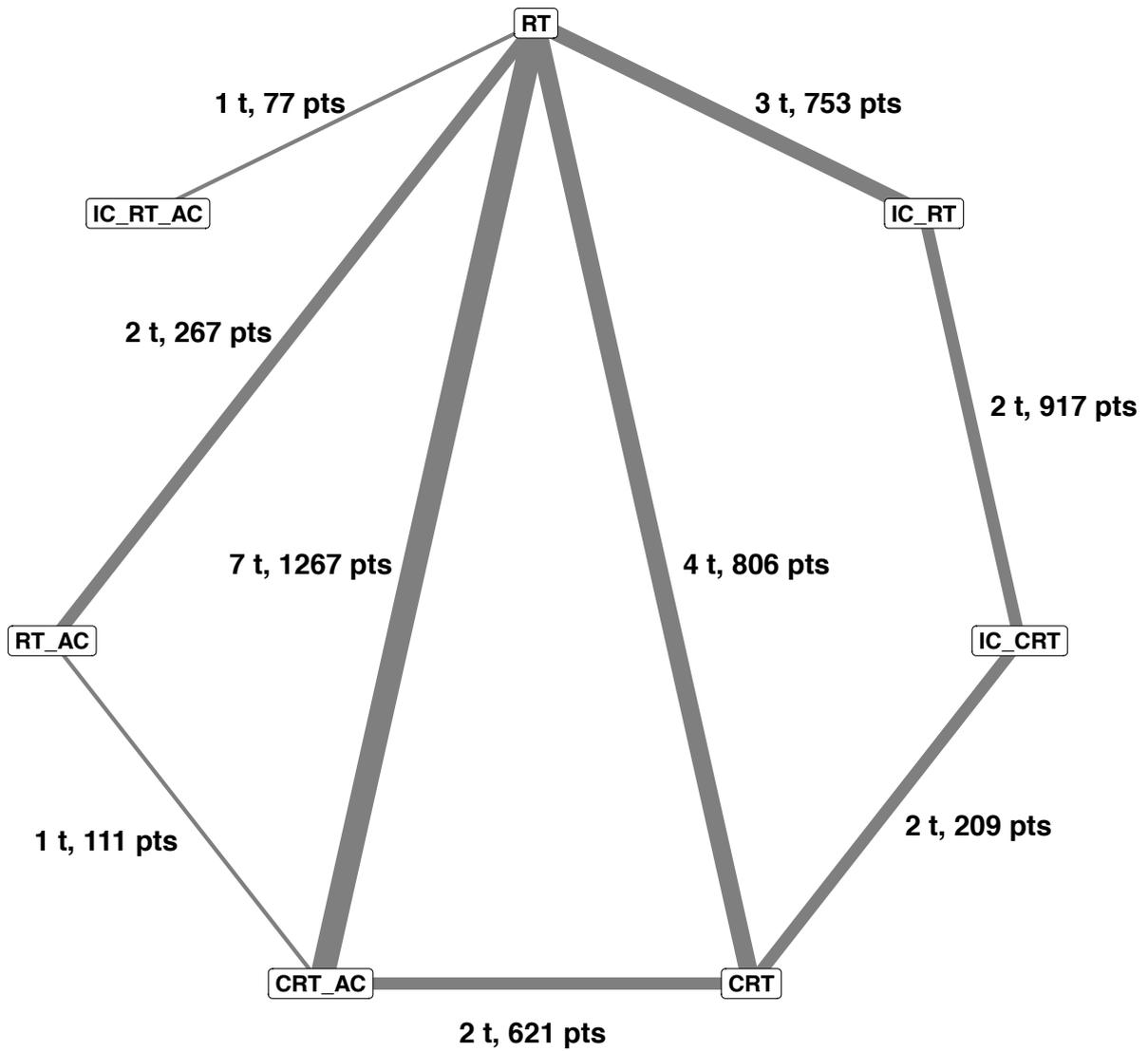

**Figure 2** Structure of the network used for the simulation study.

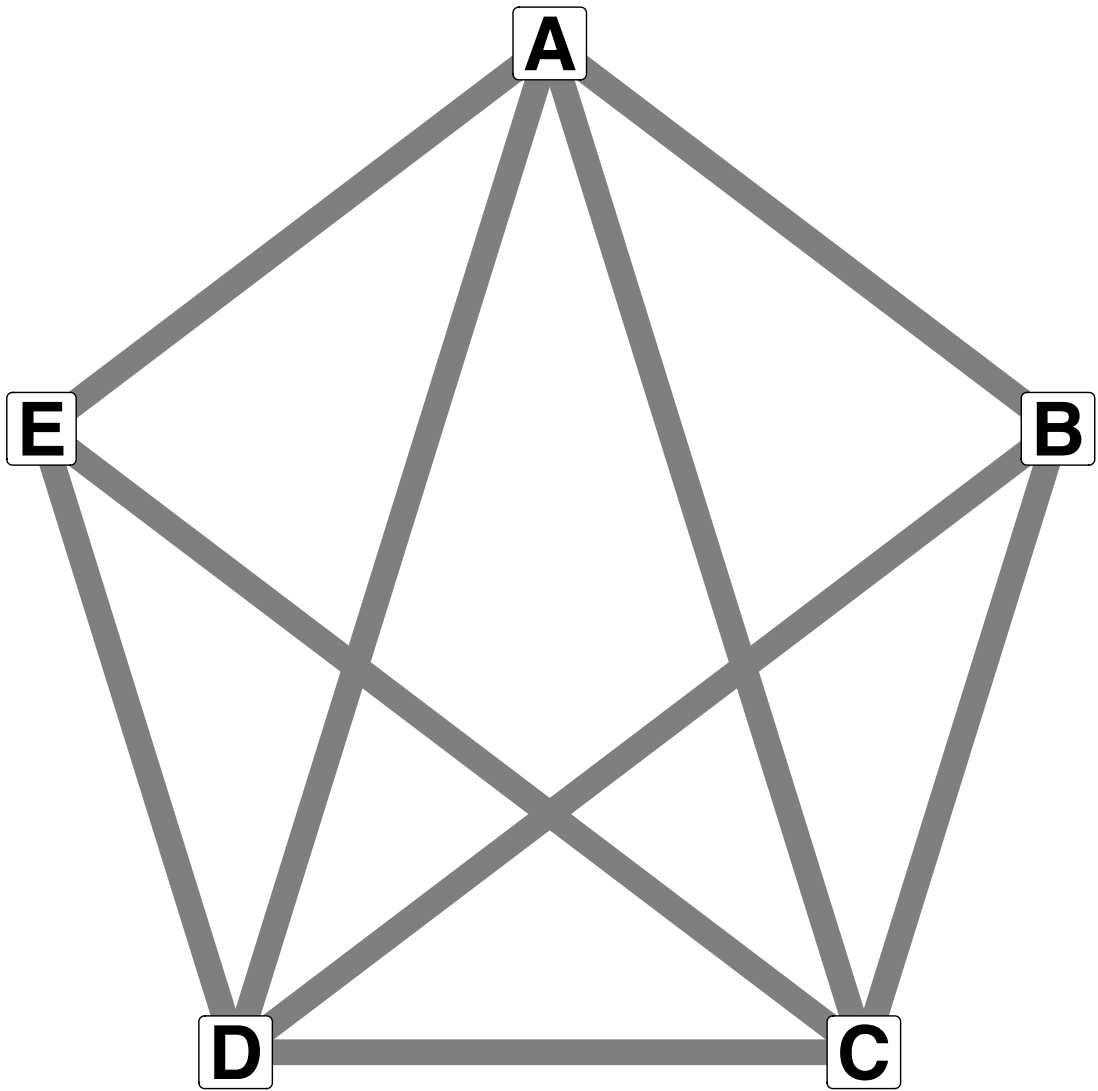

**Figure 3** Selection accuracy, False Positive Rate (FPR) and False Negative Rate (FPR) obtained in the various designs for both adaptive LASSO methods : i) adaptive LASSO method that takes into account between study heterogeneity (HetAdLASSO) and ii) adaptive LASSO method that do not take into account between study heterogeneity (FxAdLASSO). Orange lines represent the results of HetAdLASSO. Grey lines represent the results of FxAdLASSO. Solid, dashed and dotted lines correspond to simulation with 3, 5 and 10 trials per comparison, respectively.

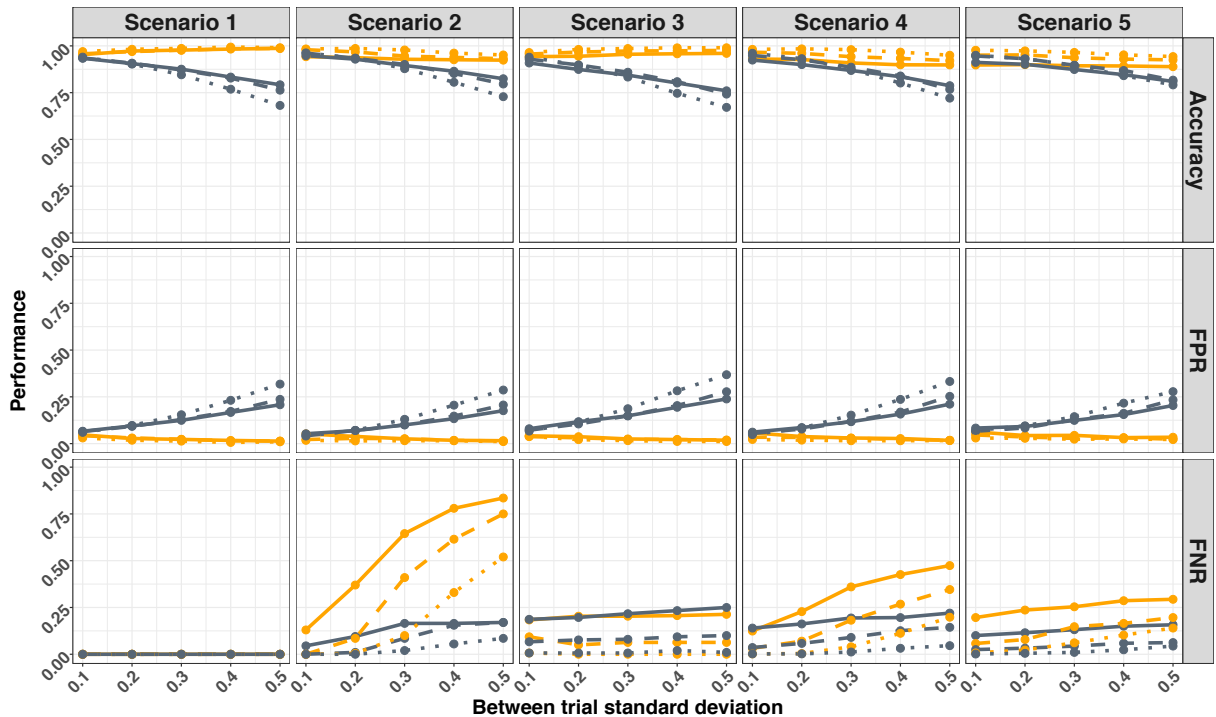

**Figure 4** Proportion of times over the 100 simulated data sets that FxAdLASSO and HetAdLASSO selected the true model. Results are stratified on parameters type: covariates, covariates-treatments interactions, inconsistency and non-proportionality. Orange lines represent the results of HetAdLASSO. Grey lines represent the results of FxAdLASSO. Solid, dashed and dotted lines correspond to simulation with 3, 5 and 10 trials per comparison, respectively.

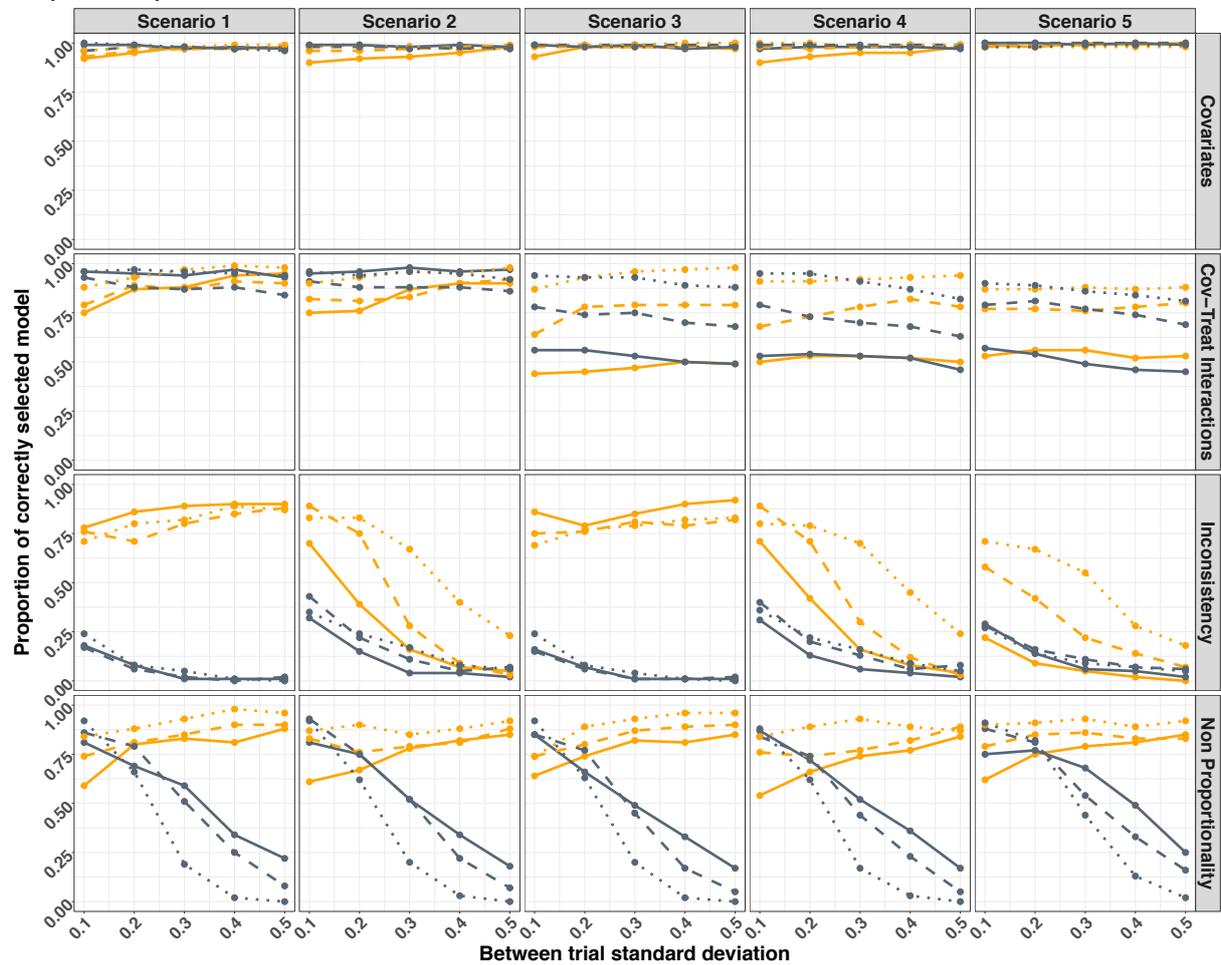

**Figure 5** Absolute bias for log-hazard ratio $\left(\bar{\beta}_q\right)_{q=1,\ldots,Q}$ parameter estimates obtained with FxAdLASSO and HetAdLASSO for simulations with 3, 5 and 10 trials per comparison. Orange boxplots represent the results of HetAdLASSO. Grey boxplots represent the results of FxAdLASSO.

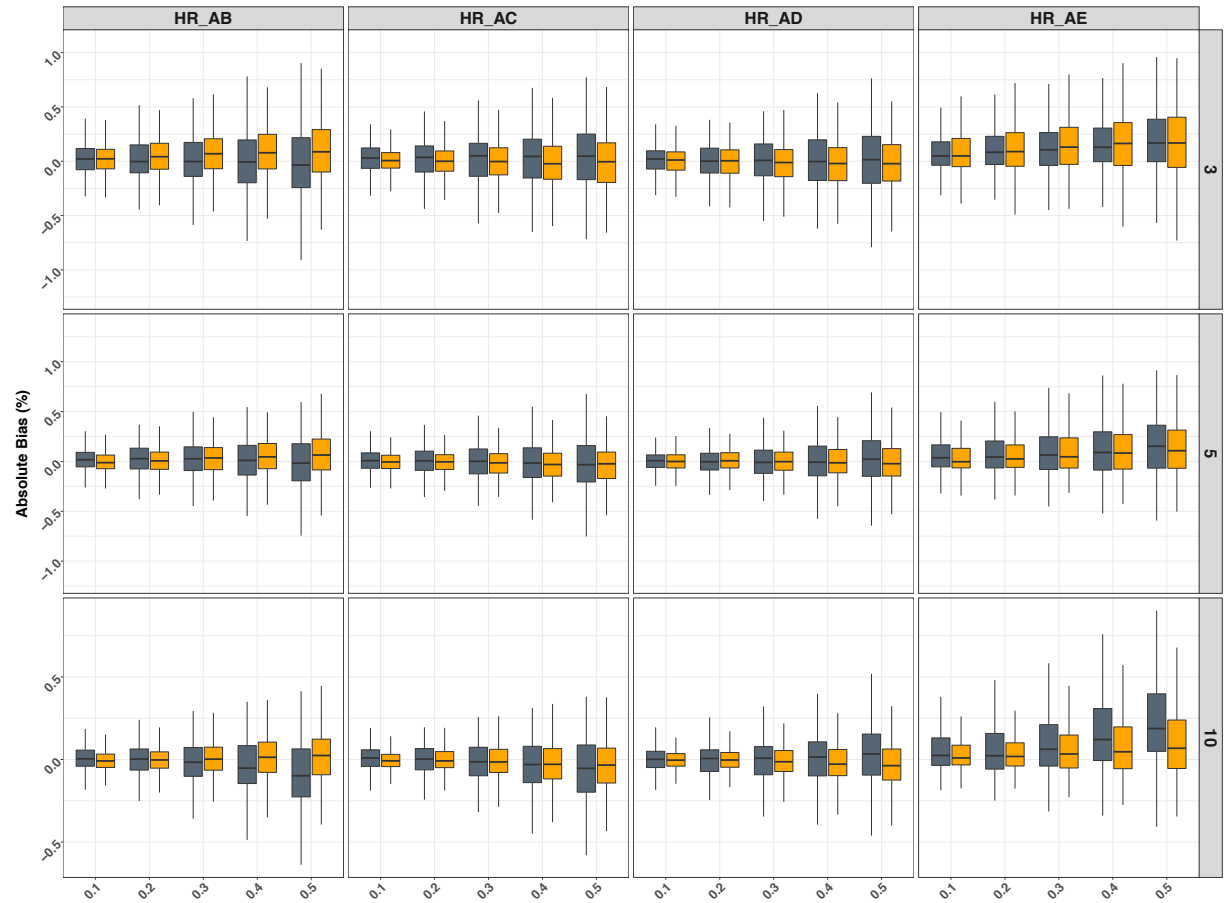

**Figure 6** Absolute bias for between-trial standard deviation ( $(\tau_q)_{q=1,\ldots,Q}$ ) estimations based on equation 4 using the optimal values of penalty parameters ( $(\lambda_q)_{q=1,\ldots,Q}$ ) for simulations with 3, 5 and 10 trials per comparison.

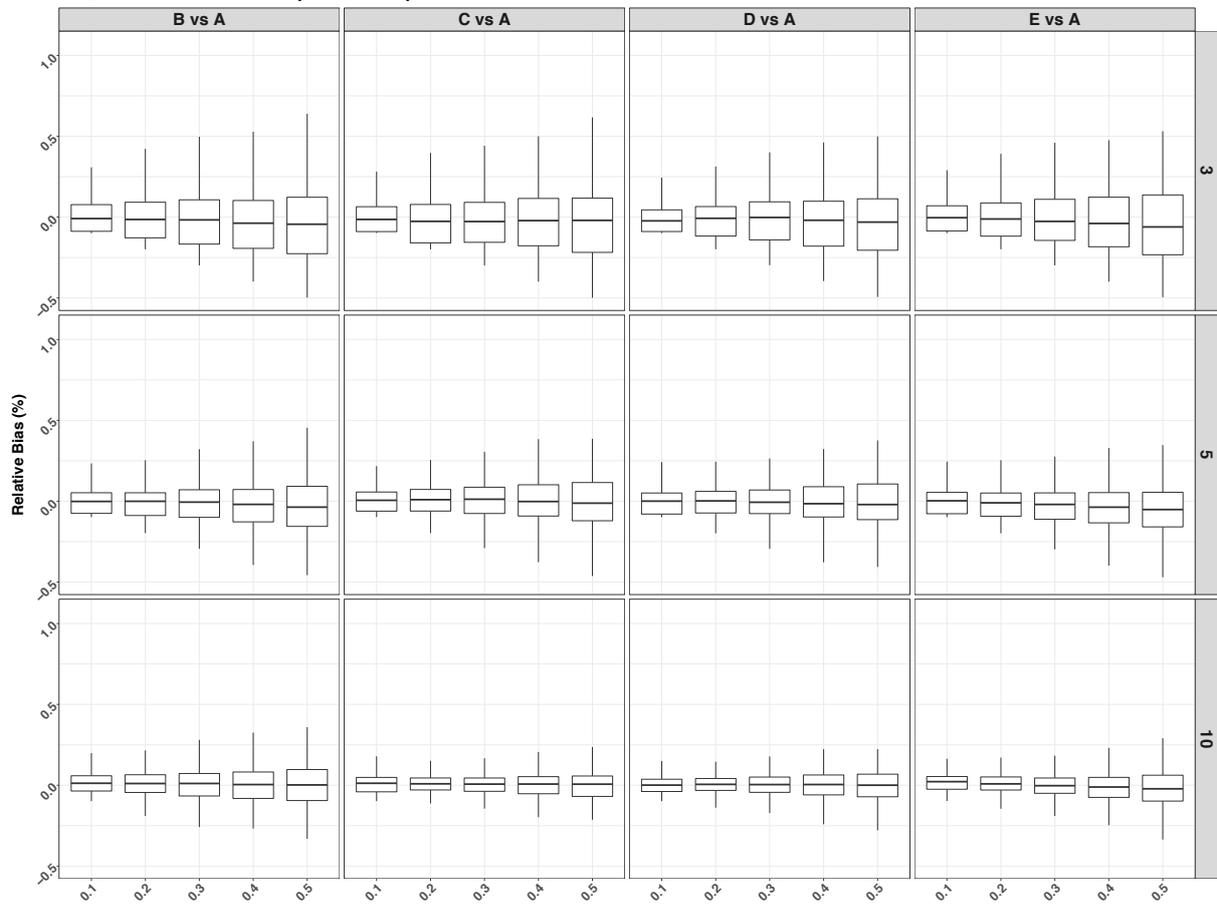

Table 1: Simulation parameters of the five scenarios

| Scenarios | Inconsistency | Covariate | Covariate-treatment interaction | Non-proportionality of risks |
|---|---|---|---|---|
| S1 | $\omega = 0$ | $\delta = 0$ | $\alpha = 0$ | $\zeta = 0$ |
| S2 | Treatments loops ABC and ADE. $\omega_{BC} = \log(0.5)$, $\omega_{DE} = \log(2)$ and 0 elsewhere | $\delta = 0$ | $\alpha = 0$ | $\zeta = 0$ |
| S3 | $\omega = 0$ | $\delta = (0, \log(1.25))$ | $\alpha_{2B} = \log(1.25)$, $\alpha_{2D} = \log(1.25)$ and 0 elsewhere. | $\zeta = 0$ |
| S4 | Treatments loops ABC and ADE. $\omega_{BC} = \log(0.5)$, $\omega_{DE} = \log(2)$ and 0 elsewhere. | $\delta = (0, \log(1.25))$ | $\alpha_{2B} = \log(1.25)$, $\alpha_{2D} = \log(1.25)$ and 0 elsewhere. | $\zeta = 0$ |
| S5 | Treatments loops ABC and ADE. $\omega_{BC} = \log(0.5)$, $\omega_{DE} = \log(2)$ and 0 elsewhere. | $\delta = (0, \log(1.25))$ | $\alpha_{2B} = \log(1.25)$, $\alpha_{2D} = \log(1.25)$ and 0 elsewhere. | $\zeta_E \neq 0$ |

Table 2: Model selected with fixed adaptive Lasso (FxAdLASSO) and random adaptive Lasso (HetAdLASSO) methods for the individual patient data network meta-analysis of chemotherapy and radiotherapy in nasopharyngeal carcinoma.

|  | **FxAdLASSO** | **HetAdLASSO** |
|---|---|---|
| **Covariates** | $\delta_{Age}, \delta_{Sex}, \delta_{T3}, \delta_{T4}, \delta_{N2}, \delta_{N3}, \delta_{IMRT}$ | $\delta_{Age}, \delta_{Sex}, \delta_{T4}, \delta_{N3}$ |
| **Covariates-by-treatment interactions** | None | None |
| **Inconsistency** | $\omega_{CRT\ CRT\_AC}$ | None |
| **Non proportionality** | None | None |

Table 3: Unpenalized re-estimations of the model selected using fixed adaptive lasso (FxAdLASSO) and random adaptive lasso (HetAdLASSO) methods. Confidence intervals were computed by bootstrapping estimates of the selected model. The column 'Two step NMA' reports the results from the two-step network meta-analysis (NMA) also used in Ribassin-Majed et al.

|  | Two step NMA | FxAdLASSO | HetAdLASSO |
|---|---|---|---|
| **Treatment effects** | | | |
| $HR_{IC\_RT\ Vs\ RT}$ | 0.92 [0.76 ; 1.13] | 0.99 [0.84 ; 1.16] | 0.94 [0.77 ; 1.16 ] |
| $HR_{IC\_CRT\ Vs\ RT}$ | 0.81 [0.63 ; 1.04] | 0.89 [0.72 ; 1.11] | 0.84 [0.63 ; 1.11 ] |
| $HR_{CRT\ Vs\ RT}$ | 0.77 [0.65 ; 0.93] | 0.75 [0.64 ; 0.87] | 0.76 [0.55 ; 1.05 ] |
| $HR_{CRT\_AC\ Vs\ RT}$ | 0.63 [0.55 ; 0.73] | 0.67 [0.59 ; 0.77] | 0.63 [0.54 ; 0.73 ] |
| $HR_{RT\_AC\ Vs\ RT}$ | 0.99 [0.75 ; 1.32] | 0.89 [0.70 ; 1.11] | 0.93 [0.69 ; 1.25 ] |
| $HR_{IC\_RT\_AC\ Vs\ RT}$ | 1.29 [0.62 ; 2.73] | 1.46 [0.83 ; 2.34] | 1.28 [0.68 ; 2.42 ] |
| **Between trial standard deviation** | | | |
| $\tau_{IC\_RT}$ | | | 0.05 |
| $\tau_{IC\_CRT}$ | | | 0.05 |
| $\tau_{CRT}$ | - | - | 0.34 |
| $\tau_{CRT\_AC}$ | | | 0.04 |
| $\tau_{RT\_AC}$ | | | 0.05 |
| **Inconsistency (Hazard ratio)** | | | |
| $\omega_{CRT\ CRT\_AC}$ | - | 0.44 [0.28 ; 0.68] | - |
| **Covariates (Hazard ratio)** | | | |
| Age (year) | - | 1.03 [1.03 ; 1.04] | 1.03 [1.02 ; 1.03] |
| Sex (Ref : Male) | - | 0.79 [0.71 ; 0.88] | 0.79 [0.71 ; 0.88] |
| T3 (Ref : T1) | - | 1.35 [1.21 ; 1.51] | - |
| T4 (Ref : T1) | - | 2.04 [1.83 ; 2.28] | 1.47 [1.33 ; 1.62] |
| N2 (Ref : N0) | - | 1.63 [1.46 ; 1.81] | - |
| N3 (Ref : N0) | - | 2.52 [2.23 ; 2.85] | 1.62 [1.46 ; 1.80] |
| IMRT (Ref : no IMRT) | - | 0.55 [0.38 ; 0.78] | - |